\documentclass[12pt]{article}
\usepackage{amsthm,amssymb,amsmath}
\usepackage[english]{babel}

\newcommand{\R}{\ensuremath{\mathbb{R}}}
\newcommand{\N}{\ensuremath{\mathbb{N}}}

\newcommand{\bx}{{\boldsymbol{x}}}

\renewcommand{\d}{\operatorname{d}}
\newcommand{\I}{\operatorname{i}}
\newcommand{\e}{\operatorname{e}}

\newcommand{\be}{\begin{equation}}
\newcommand{\ee}{\end{equation}}
\begin{document}

\title{\sc A hodograph transformation
which applies to the heavenly equation\thanks{Partially supported
by CICYT proyecto PB98--0821 }}
\author{Manuel Ma{\~n}as$^{1,\dag}$ and Luis Mart{\'\i}nez Alonso$^{1,\ddag}$
\\
\emph{ $^1$Departamento de F\'{\i}sica Te\'{o}rica II, Universidad
Complutense}\\ \emph{E28040 Madrid, Spain} \\
\texttt{$^\dag$manuel@darboux.fis.ucm.es}\\
\texttt{$^\ddag$luism@fis.ucm.es}}
\date{} \maketitle
\begin{abstract}
A hodograph transformation for a wide family of multidimensional
nonlinear partial differential equations is presented. It is used
to derive  solutions of the heavenly equation (dispersionless Toda
equation) as well as a family of explicit ultra-hyperbolic
selfdual vacuum spaces admiting only one Killing vector which is
not selfdual, we also give the corresponding explicit
Einstein--Weyl structures.
\end{abstract}

\vspace*{.5cm}

\begin{center}\begin{minipage}{12cm}
\emph{Key words:} Heavenly equation, hodograph transformations,
Einstein--Weyl spaces.

\emph{ 1991 MSC:} 58B20.
\end{minipage}
\end{center}
\newpage

\section{Introduction}

This work introduces a hodograph method to construct solutions of
a ample family of nonlinear partial differential equations (PDE)
among which we  have the dispersionless Kadomtsev-Petviashivili
(dKP) equation and the heavenly equation,  relevant in the finding
of Einstein--Weyl 3D spaces and selfdual vacuum Einstein spaces
\cite{boyer-finley,dunajski,1}. Hodograph transformations goes
back the XIX century and as was shown by Riemann they are relevant
in the discussion of hydrodynamic type systems, this hodograph
transformation was generalized recently by Tsarev \cite{tsarev}.

The layout of this paper is as follows. The next section is
devoted to describe our scheme. Using solutions to a implicit
relation we find solutions to nonlinear PDEs. Finally, in the \S 3
we apply these results to the finding of new solutions of the
heavenly equation and of Einstein--Weyl spaces and the
corresponding ultra-hyperbolic self-dual vacuum Einstein spaces.
At this point is important to mention that our solutions are
different from those found in \cite{1}.

\section{The hodograph transformation}

Our method begun with the following implicit equation for
determining a scalar function $u=u(\bx)$ depending on $n$
variables $\bx=(x_1,\ldots,x_n)$
\begin{equation}\label{1.1}
X_0(u)+\sum_{i=1}^n x_i\, X_i(u)=0,\
\end{equation}
where $X_i$, $i=0,1,\dots,n$, are given functions of $u$. By
denoting $x=x_1,\; t_i=x_{i+1},\; (i=1,\ldots,n-1)$, it follows
that \eqref{1.1} is a hodograph transformation for the family of
one-dimensional hydrodynamical systems
\begin{equation}\label{1.2}
u_{t_i}=C_i(u)u_x,\quad i=1,\ldots,n-1,
\end{equation}
where
\begin{equation}
C_i(u):=\frac{X_{i+1}(u)}{X_1(u)}.
\end{equation}

Our main observation is that \eqref{1.1} provides solutions for
the family of nonlinear PDE´s
\begin{equation}\label{1.3}
\sum_{|\alpha|=m} c_{\alpha}D^{\alpha}\phi=D^{\beta}
F(\phi),\quad |\beta|=m,
\end{equation}
where $D^{\alpha}$ and $D^{\beta}$ denote partial differentiation
operations of a given order $m$ corresponding to $n$-component
multi-indices $\alpha,\beta\in\N^n$, $F=F(\phi)$ is an arbitrary
function and $c_{\alpha}$ are arbitrary constants. We are going
to prove that a solution of \eqref{1.3} is given by the function
\begin{equation}  \label{phiGQ}
\phi(\bx):=G(Q(u)),\; Q(u):= \frac{\sum_{|\alpha|=m}c_{\alpha}
X^{\alpha}(u)}{X^{\beta}(u)},\quad X^{\sigma}:=X_{\sigma_1}\cdots
X_{\sigma_n},
\end{equation}
where $G:=(F_u)^{-1}$ is the inverse function of the derivative
$F_u$ of $F$ with respect to $u$.

From \eqref{phiGQ}  we deduce that
\begin{align*}
\phi_{x_i}&=G'(Q(u))Q'(u)u_{x_i}=
G'(Q(u))Q'(u)\frac{X_i(u)}{X_j(u)}u_{x_j}\\&=
\frac{\partial}{\partial
x_j}\int^{u}G'(Q(u))Q'(u)\frac{X_i(u)}{X_j(u)}\d u
\end{align*}
and therefore
\[
D^\alpha\phi=D^\beta\int^{u}G'(Q(u))Q'(u)
\frac{X^\alpha(u)}{X^\beta(u)}\d u.
\]
From this relation we conclude
\begin{align*}
\sum_{|\alpha|=m}c_{\alpha} D^\alpha\phi&=
  D^\beta\int^{u}G'(Q(u))Q'(u)\frac{
  \sum_{|\alpha|=m}c_{\alpha}X^\alpha(u)}{X^\beta(u)}\d u
 \\&=D^\beta \int^{u}G'(Q(u))Q'(u) Q(u)\d u,
\end{align*}
Now, if $H:=F\circ G$ then
\[
(H)'(Q)=(F'\circ G)(Q) G'(Q)=Q G'(Q)
\]
 and hence
\[
\sum_{|\alpha|=m}c_{\alpha} D^\alpha\phi=
  D^\beta\int^{u}H'(Q(u))Q'(u)\d u
 =D^\beta H(Q)=D^\beta F(\phi).
\]
\paragraph{Observations}
\begin{enumerate}
\item In spite of the implicit nature of the hodograph relation
\eqref{1.1} we can easily find  explicit examples. We shall point
out two of such cases:
\begin{itemize}
\item Assuming that
\[
X_i(u)=\sum_{j=0}^N a_{ik}u^k,\quad i=0,1,\dots,n,
\]
\eqref{1.1} reads as
\[
\sum_{k=0}^NA_k u^k=0,\quad A_k=a_{0k}+\sum_{i=1}^na_{ik}x_i,
\]
and we will have $N$ complex roots
\[
u_l=u_{l}(A_0,\dots,A_N),\quad l=1,\dots,N.
\]

For each of these roots we can evaluate $X_i(u_l)$ and get a
family of solutions. As we know if $N\leq 4$ the roots can be
gotten explicitly and therefore we  will have an explicit
algebraic function depending on the parameters
$\{a_{ik}\}_{\substack{k=0,\dots,4,\\i=1,\dots,n}}$. 
\item Another example appears by considering the Lambert
function $W(z)$ which solves
\[
W\exp(W)=z
\]
and has been studied with certain detail \cite{lambert}. The point
here is that the implicit \emph{fundamental}  relation
\[
a+b u+c\exp(u)=0
\]
is solved in terms of the Lambert function as follows
\[
u=-W\Big(\frac{c}{b}\exp\Big(-\frac{a}{b}\Big)\Big)-\frac{a}{b}.
\]
Thus, taking
\[
X_i(u)=a_i+b_i u+c_i\exp(u),
\]
with $a_i,b_i$ and $c_i$ arbitrary constants, the hodograph
relation is
\[
A+B u+C\exp(u)=0,
\]
with
\[
A:=a_0+\sum_{i=1}^na_ix_i,\,B:=b_0+\sum_{i=1}^nb_ix_i,\,C:=c_0+\sum_{i=1}^nc_ix_i,
\]
and the solution is
\[
u=-W\Big(\frac{C}{B}\exp\Big(-\frac{A}{B}\Big)\Big)-\frac{A}{B}.
\]
Hence, introducing the rational functions
\[
r(\bx):=\frac{a_0+\sum_{i=1}^na_ix_i}{b_0+\sum_{i=1}^nb_ix_i},\quad
s(\bx):=\frac{b_0+\sum_{i=1}^nb_ix_i}{c_0+\sum_{i=1}^nc_ix_i}
\]
 we can evaluate
\[
X_i(\bx)=a_i-b_ir(\bx)-\big(b_iC-c_is(\bx)\big)
W\Big(\frac{1}{s(\bx)\exp r(\bx)}\Big)
\]
and using \eqref{phiGQ} get a solution to the nonlinear PDE
\eqref{1.3} in terms of the Lambert function.
\end{itemize}
\item We can employ the freedom in the the choice for the
functions $\{X_i(u)\}_{i=1}^n$ to generate solutions of more
general equations. Suppose  a functional dependence of the form
\[
\phi=(F_\alpha')^{-1}(Q_\alpha)=(F_\beta')^{-1}(Q_\beta),
\]
for  all $\alpha,\beta\in\mathcal{I}$, being $\mathcal{I}$ a set
of $r=\operatorname{card}\mathcal{I}$ multi-indices of order $m$,
and
\[
Q_\gamma X^\gamma=\sum_{|\delta|=m}a_\delta X ^\delta
\]
 Then, $\phi$ satisfies
\[
\sum_{|\delta|=m}a_\delta
D^\delta\phi=\frac{1}{r}\sum_{\gamma\in\mathcal{I}}D^\gamma
F_\gamma(\phi).
\]
For example, the hodograph relation
\[
tT(u)+xX(u)+yY(u)=H(u)
\]
provides solutions to
\[
\frac{1}{2}(\phi_{xx}+\phi_{yy})=(\exp(\phi))_{tt},\quad
\phi=\log\frac{X^2+Y^2}{2T^2}
\]
as well to
\[
\frac{1}{2}(\phi_{xx}+\phi_{yy})=(\exp(2\phi))_{xt},\quad
\phi=\log\sqrt{\frac{X^2+Y^2}{4XT}}.
\]
Thus, we need to fulfill
\[
X^3+Y^2X=T^3.
\]
So that, the solutions of
\[
t\sqrt[3]{X(u)+Y^2(u)/X(u)}+xX(u)+yY(u)=H(u)
\]
gives
\[
\phi_{xx}+\phi_{yy}=(\exp(\phi))_{tt}+(\exp(2\phi))_{xt}, \quad
\phi=\frac{1}{3}\log(1+(Y/X)^2)-\log(2).
\]
\end{enumerate}

\section{Applications in General Relativity}

Among the nonlinear PDEs of the form for which our hodograph
technique is applicable one finds an integrable equation: the dKP
equation
\[
\phi_{tx}+\phi_{yy}=(\phi^2)_{xx}.
\]
This equation is relevant in hydrodynamics and  our hodograph
solutions were already discussed by Kodama in \cite{kodama}, the
dKP equation appears in the construction of three-dimensional
Einstein--Weyl spaces \cite{dunajski}. Another integrable equation
within our family of PDEs is known with different names: heavenly
equation, Boyer--Finley equation, dispersionless Toda and
SU$(\infty)$-Toda equation:
\begin{equation}\label{he}
\phi_{z\bar z}+\kappa(\e^\phi)_{tt}=0,\quad\kappa =\pm 1.
\end{equation}
where $z=x+\I y$ and $\bar z=x-\I y$, $x,y,t,\phi\in\R$. This
equation has been found to characterize  self-dual vacuum Einstein
spaces ---of signature $(++--)$ (ultra-hyperbolic) for $\kappa
=-1$ and $(++++)$ (Euclidean) when $\kappa =1$---
 having a non-selfdual  Killing vector  \cite{boyer-finley},
 while those having a
selfdual Killing vector appear to be related to the wave (or
Laplace) equation and the metrics are of Gibbons--Hawking type
\cite{gibbons-hawking}.

Very few solutions of the heavenly equation have been found. In
first place a separation of variables $\phi(z,\bar z,t) =
\log(f(t)) + \Phi(z,\bar z)$ leads to  the Liouville equation
\cite{liouville} $\Phi_{z\bar z}=\e^\Phi$, whose general solution
is well known. If one imposes a symmetry, say $z=\bar z$, then the
equation linearize, after a hodographic change of variable
\cite{1} and in this form implicit solutions are gotten. Also in
\cite{tod-painleve} an implicit solution based on the Painleve
equations was given. In \cite{tod-calderbank} a new explicit
solution was presented, see also \cite{invariantes,nutku}. Further
studies of the geometry associated with the equation can be found
in for example \cite{calderbank-gt}. See also
\cite{saveliev,strachan} for further information regarding this
equation.

The heavenly equation is also known as the dispersionless Toda
equation and appears as an example of the so called Whitham
hierarchies. It has been applied to the study of conformal
transformations and topological field theory \cite{toda}.

Our scheme provides  solutions to the ultra-hyperbolic heavenly
equation. The problem is to find solutions of the heavenly
equation so that the corresponding metric does not have an
additional Killing vector. Hence, following \cite{nutku} the
solutions of the heavenly equation must be non-invariant
\cite{invariantes} (being the symmetry group composed of
translations, scaling and conformal transformations), as these
symmetries will carry to corresponding additional Killing vectors.
This construction is  equivalent to self-dual hyper-K{\"a}hler spaces
and, as was shown by Ward \cite{1} the heavenly equation can be
used to generate Einstein--Weyl spaces in 3D.

To check that our scheme gives solutions of non-invariant type,
for the ultra-hyperbolic case, we shall use the hodograph equation
in the following form
\[
t+\rho\e^{-\I\alpha(\rho)}z+\rho\e^{\I\alpha(\rho)}\bar z
=h(\rho),
\]
where $\alpha$ and $h$ are arbitrary functions of $u=\rho$ and the
solution of the heavenly equation is given by
\[
\phi=\log(\rho^2),
\]
this form of the hodograph equation ensures that $\phi$ takes real
values. Using polar coordinates $z=r\e^{\I\theta}$ we get the
following hodograph relation
\begin{equation}\label{hodografa-coseno}
t+2\rho r\cos(\alpha(\rho)-\theta)=h(\rho).
\end{equation}

Now, following \cite{invariantes} we must check whether or not is
possible to find constants $\alpha$ and $\beta$ and functions
$a(z)$ and $b(\bar z)$ such that the following equation holds
\begin{equation}\label{invariant-solutions}
(\alpha+\beta t)\phi_t+a(z)\phi_{z}+b(\bar z)\phi_{\bar z
}=2\beta-a'(z)-b'(\bar z).
\end{equation}
Now, recalling that the hodograph relation implies
\begin{align*}
\rho_z=
\frac{\rho\e^{-\I\alpha}}{D},\quad
\rho_{\bar z}=\frac{\rho\e^{\I\alpha}}{D},\quad
\rho_t=\frac{1}{D},
\end{align*}
with
\[
D:=h'-(1-\I\rho\alpha')\e^{-\I\alpha}z-
(1+\I\rho\alpha')\e^{\I\alpha}\bar z,
\]
 and introducing the notation
\[
A(z):=a(z)-\beta z,\quad B(\bar z):=b(\bar z)-\beta\bar z
\]
we can write  \eqref{invariant-solutions} in the following form
\begin{equation}\label{invariant-solutions2}
\alpha+\beta h+ A\rho\e^{-\I\alpha}+ B\rho\e^{\I\alpha}=
-(A'+B')F,
\end{equation}
with
\[
F:= \frac{\rho D}{2}.
\] Now, if the functions $\{1,\rho\e^{-\I\alpha},\rho\e^{\I\alpha},h\}$
 are linearly dependent,
\begin{equation}\label{linear-dependence}
\lambda_1 1+\lambda_2 \rho\e^{-\I\alpha}+\lambda_3
\rho\e^{\I\alpha}+\lambda_4 h=0,
\end{equation}
for some constants $\lambda_i$, $i=1,2,3,4$,
 then \eqref{invariant-solutions2} will be
identically satisfied if
$\alpha=\lambda_1,\beta=\lambda_2,A=\lambda_3$ and $B=\lambda_4$.
 The invariant solutions should appear also if
\eqref{invariant-solutions2} holds taking $x,y,t$ and $u$ as
independent variables. In doing so must impose $A=A_1z+A_0$ and
$B=B_1\bar z+B_0$ together with the equations
\begin{align*}
A_1-\frac{1}{2}(1-\I\rho\alpha')(A_1+B_1)&=0,\\
B_1-\frac{1}{2}(1+\I\rho\alpha')(A_1+B_1)&=0,\\
\alpha+\beta h+ A_0\rho\e^{-\I\alpha}+
B_0\rho\e^{\I\alpha}+\frac{1}{2}(A_1+B_1)\rho h'&=0.
\end{align*}
The two first are equivalent to the ODE
\[
\alpha'=\I\frac{A_1-B_1}{A_1+B_1}\frac{1}{\rho}
\]
that implies
\begin{equation}\label{eq-f}
\alpha=\I\gamma\log\rho+C\Rightarrow
\e^{-\I\alpha}=c\rho^\gamma,\quad \gamma:=\frac{A_1-B_1}{A_1+B_1}.
\end{equation}
while the third determines $h$ as a solution of the following ODE
\[
\alpha+\beta h + A_0c\rho^{1+\gamma}+
B_0c^{-1}\rho^{1-\gamma}+\frac{1}{2}(A_1+B_1)\rho h'=0,
\]
whose solution is
\begin{equation}\label{eq-g}
h(\rho)=C\rho^{2\frac{\beta}{A_1+B_1}}-\frac{\alpha}{\beta}-
\frac{cA_0}{A_1+B_1}\rho^{2\frac{A_1}{A_1+B_1}}-
\frac{B_0}{c(A_1+B_1)}\rho^{2\frac{B_1}{A_1+B_1}}.
\end{equation}

 Generically, if neither \eqref{linear-dependence} nor
 \eqref{eq-f}-\eqref{eq-g} hold it would be difficult to
 have an invariant solution. Introducing the notation
$f_{\pm}(\rho):=\rho\e^{\mp\I\alpha}$, $F_\pm:=f_\pm'/F'$ and
$H:=h'/F'$ taking taking $t$-derivatives
 of \eqref{invariant-solutions2} we get
 \begin{align}
\beta H+AF_++B F_-&=-(A'+B'),\\
\label{der-inv2}
 \beta H^{(n)}+AF_+^{(n)}+B F_-^{(n)}&=0,\quad n\geq 1.
\end{align}
Thus, in order to have invariant solutions we must impose
\[
\begin{vmatrix} H^{(n_1)} & F_+^{(n_1)}&F_-^{(n_1)}\\
H^{(n_2)} & F_+^{(n_2)}&F_-^{(n_2)}\\
H^{(n_3)} & F_+^{(n_3)}&F_-^{(n_3)}\end{vmatrix}=0, \text{with }
0<n_1<n_2<n_3,\quad n_j\in\N.
\]
 and therefore an infinite
set of equations need to be satisfied by the solution $\rho(z,\bar
z,t)$ of the hodograph relation.


Following \cite{1} we know that any solution of $\phi$ of
\eqref{he} defines an Einstein--Weyl space given by
\begin{equation}\label{2.2}
\d l^2=\d t^2-4 \rho^2(\d r^2+r^2\d\theta^2), \quad \omega=2\phi_t
\d t.
\end{equation}

Thus, by introducing the change of variables
$(t,r,\theta)\rightarrow (\rho,r,\psi)$
\begin{equation}\label{2.5}
\begin{aligned}
t&=h(\rho)-2r\rho\cos\psi,\\
r&=r,\\
\theta&=\psi+\alpha(\rho),
\end{aligned}
\end{equation}
the corresponding  Einstein--Weyl structure becomes explicitly
given in terms of two arbitrary functions $\alpha(\rho)$ and
$h(\rho)$  by
\begin{align*}\nonumber
\d l^2&=[(h'-2r\cos\psi)^2-4r^2\rho^2(\alpha')^2]\d \rho^2
-4\rho^2\sin^2 \psi \d r^2-4r^2\rho^2\cos^2 \psi \d \psi^2\\
&
-2\rho\cos(\psi)(h'-2r\cos\psi)\d\rho\d r- 4r\rho[(h'-2r\cos\psi)\sin(\psi)-2r\rho\alpha']\d \rho\d \psi\\
&-2r \rho^2\sin 2\psi\d r\d \psi,
\end{align*}
and
\begin{equation}\label{2.6}
\omega=\frac{4}{\rho}\frac{(h'-2r\cos\psi)\d\rho -2\rho\cos\psi\d
r+2r\rho\sin\psi\d\psi}{h'-2r\cos\psi-2r\rho\alpha'\sin \psi}\d t.
\end{equation}
It should be noticed that \eqref{2.5}--\eqref{2.6} define a family
of Einstein--Weyl structures different from that characterized by
Ward in \cite{1}. Indeed, Ward uses an hodograph transformation
for determining all solutions of \eqref{he} independent on one of
the spatial variables $x$ or $y$.

The corresponding ultra-hyperbolic vacuum Einstein metric in 4D is
given by \cite{nutku}
\begin{align*}
\d s^2&=\phi_t\d l^2-\frac{1}{\phi_t}[\d \tilde t+\I(\phi_z\d
z-\phi_{\bar z}\d\bar z)]^2\\&= \frac{2}{\rho D}\d l^2- \frac{\rho
D}{2}\big(\d \tilde t-\frac{4}{D}(\sin\psi\d r+r\cos\psi\d
\psi+\alpha'r\cos\psi\d\rho)\big)^2,
\end{align*}
with $D=h'-2r\cos\psi-2r\rho\alpha'\sin \psi$.

\section{Acknowledgements}
The authors would like to acknowledge discussions with Alexander
Mikhailov, Ian Strachan and Sergei Tsarev.

\end{document}